\documentclass[10pt,prl,aps,twocolumn]{revtex4-1}

\usepackage{graphicx,bm}
\usepackage{amsmath}
\usepackage{amssymb}
\usepackage{epsfig}
\usepackage{epsf}
\usepackage{verbatim}
\usepackage{hyperref}
\usepackage[usenames]{color}

\newcommand{\beqa}{\begin{eqnarray}}
\newcommand{\eeqa}{\end{eqnarray}}
\newcommand{\beq}{\begin{equation}}
\newcommand{\eeq}{\end{equation}}

\newcommand{\spp}{S^{+}}
\newcommand{\smm}{S^{-}}

\newcommand{\mbf}[1]{\mathbf{#1}}

\begin{document}
\title{Magnon Spin Hall Magnetoresistance of a Gapped Quantum Paramagnet}
\author{Camilo Ulloa$^{1}$}
\email{C.Ulloa@uu.nl}
\author{R. A. Duine$^{1,2}$}
\affiliation{${}^1$Institute for Theoretical Physics, Utrecht University,
Princetonplein 5, 3584 CC Utrecht, The Netherlands}
\affiliation{${}^2$Department of Applied Physics, Eindhoven University of Technology,
P.O. Box 513, 5600 MB Eindhoven, The Netherlands}

\begin{abstract}
Motivated by recent experimental work we consider spin transport between a normal metal and a gapped quantum paramagnet. We model the latter as the magnonic Mott-insulating phase of an easy-plane ferromagnetic insulator. We evaluate the spin current mediated by the interface exchange coupling between the ferromagnet and the adjacent normal metal. For the strongly-interacting magnons that we consider, this spin current gives rise to a spin-Hall magnetoresistance that strongly depends on the magnitude of the magnetic field, rather than its direction. This work may motivate electrical detection of the phases of quantum magnets and the incorporation of such materials into spintronic devices.
\end{abstract}

\maketitle

\textit{Introduction.---} 
Spin transport through magnetic insulators and its actuation and detection via adjacent normal metals have been attracting a great deal of attention from the spintronics community. These developments yield the possibility to transport spin angular momentum without an accompanying charge current and thus without Joule heating. In addition to raising scientific interest this opens the possibility to use magnetic insulators to transport information, with the long-term goal of replacing electronics with a more energy-efficient solution. 

There are several experimental manifestations of the coupling, across an interface, between the magnetic order in the insulating ferromagnet (FM) with the electron spins in the normal metal (NM). The first class of experiments involves static magnetic order. Here, a prime example is spin-Hall magnetoresistance. This is the observation that the resistance of a heavy normal metal, typically Pt, depends on the relative orientation of the current and the magnetization direction of an adjacent magnetic insulator --- typically Yttrium-Iron-Garnet (YIG) \cite{Nakayama2013}. The second class of experiments involves coherent dynamics of the magnetic order, e.g., in response to a microwave field. This leads e.g. to pumping of spin current from the ferromagnet into the normal metal \cite{Ando2010,Vlietstra2013} or vice versa \cite{hamadeh2014}. The final class of experiments involves incoherent dynamics of the magnetic insulator \cite{footnote1}. In most experiments with magnetic insulators this means that the magnetic dynamics is described in terms of magnons --- quantized spin waves of the magnetic order parameter. The spin Seebeck effect, where a magnon spin current is induced by a thermal gradient and detected via the inverse spin Hall effect in an adjacent normal metal \cite{Uchida2010,Adachi2011,Wu2015}, belongs to this final class of experiments and recently has been used to probe short-ranged order in classical spin liquids \cite{Liu2017}. Another typical experiment involving incoherent magnons is a non-local transport measurement that involves two normal metals on top of a magnetic insulator. Here, magnon spin current gives rise to a non-local resistance via the spin-Hall and inverse spin-Hall effects at injector and detector, respectively   \cite{Ludo2016Nat,  Goennenwein2015, Li2016,Wesenberg2017,Thiery2017}. In a variation on this non-local experiment, Giles {\it et al.} \cite{Giles2015} injected the spin current by heating the Pt injector with a laser. 

\begin{figure}[htbp] 
	\centering
	\includegraphics[width=0.45\textwidth]{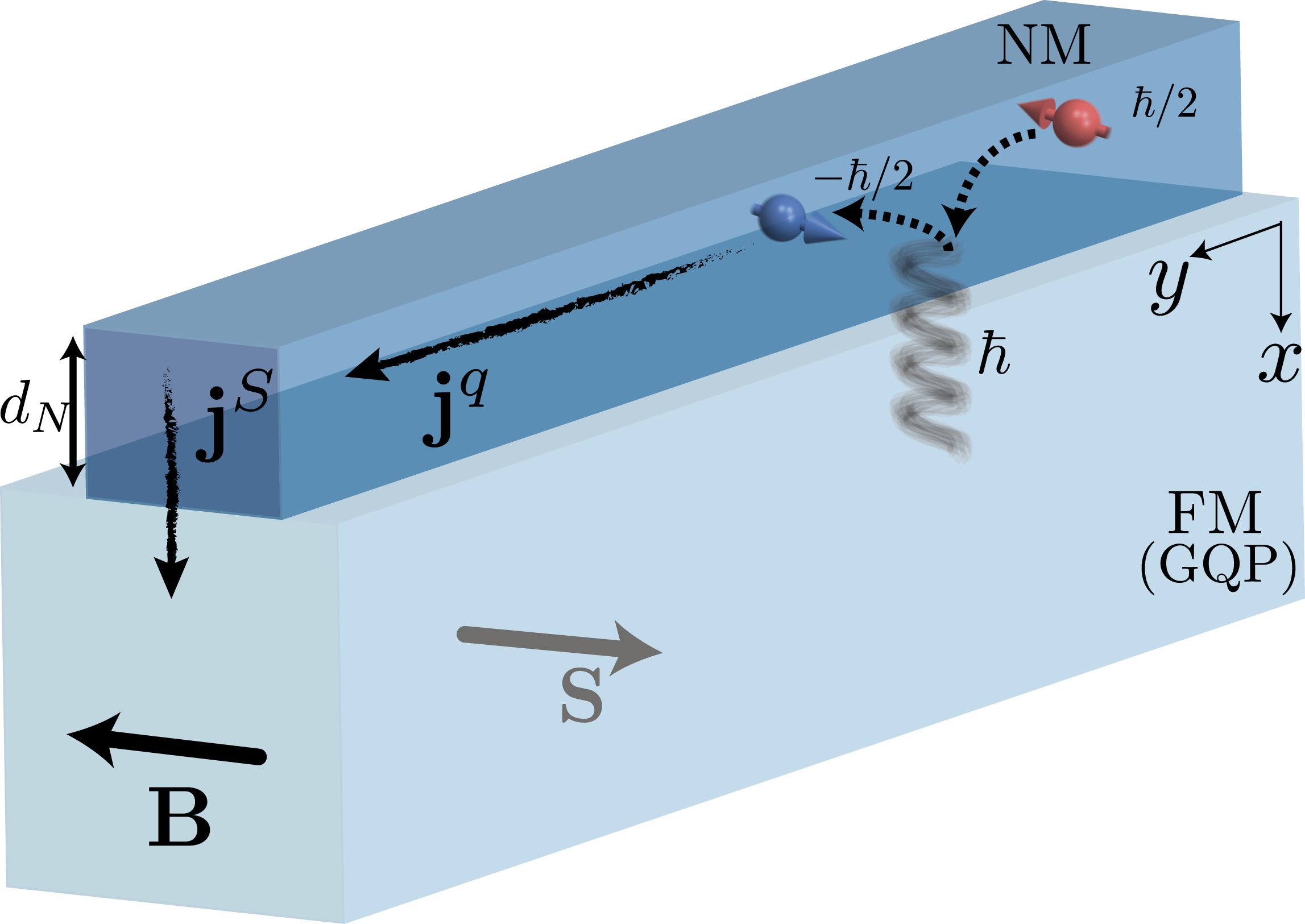} 
	\caption{(Color online.) Schematic of the NM$|$GQP heterostructure. The field $\mathbf{B}$ is taken in the $z$-direction such that the $-z$-direction is the equilibrium axis of the spins $\mathbf{S}$. In this way, magnons in the magnetic insulator --- indicated by the wavy line --- carry $\hbar$ angular momentum and are excited by electron spin flips at the interface. The spin current ($\mathbf{j}^{S}$) that arises via the spin Hall effect is spin-polarized in the $z$-direction and flows in the $x$-direction. The charge current ($\mathbf{j}^{q}$) that drives the spin Hall effect flows in the $y$ direction. The thickness of the NM is $d_N$. }
	\label{fig:device}
\end{figure}

In most theoretical studies, the coherent dynamics of the magnetization is described using the Landau-Lifshitz-Gilbert (LLG) equation \cite{Gilbert1955, LandauBook}. The incoherent dynamics is then typically incorporated using stochastic extensions of this equation \cite{Xiao2010, Hoffman2013} or by means of the Boltzmann equation \cite{Zhang2012, Rezende2014, Ludo2016}. These ways of treating the incoherent dynamics are appropriate for weakly-interacting magnons. 

The field of quantum magnetism deals with strongly-interacting magnetic materials that are out of the scope of the above-mentioned theoretical treatments \cite{Zapf2014_rev}. Some of these materials have properties, such as large heat conductivities \cite{hess2007}, that make them interesting from the point-of-view of e.g. a spin-Seebeck measurement. Advances in this direction have been made by Hirobe {\it et al.} \cite{Hirobe2016} who experimentally studied the spin Seebeck effect in a heterostructure of Pt and Sr$_2$CuO$_3$. In the interpretation of this experiment, the spin current is carried by spinons rather than magnons. 
Among the most ubiquitous quantum-magnetic systems are gapped quantum paramagnets (GQPs) \cite{Zapf2014_rev}. These are magnetic systems that, as a result of strong correlations exhibit plateaus in the magnetization as a function of applied field in part of their temperature-field phase diagram. Often, gapped quantum paramagnets are easy-plane magnetic insulators in which long-range ordering favored by exchange interactions  --- which would correspond to a spin-superfluid state --- is prevented by strong anisotropy. In this Letter we study the injection of spin current into a gapped quantum paramagnet.

This Letter is motivated in a broad sense by i) the work by Hirobe \emph{et al.}~\cite{Hirobe2016}, ii) the recent developments in magnon spintronics and quantum magnetism in general, and iii) the scientific need for a simple FM$|$NM model system in which the description of the magnetic insulator falls outside the Landau-Lifshitz-Gilbert paradigm of weakly-interacting magnons. (For work in this latter direction that considers biasing by magnetic fields rather than metallic reservoirs, see Ref.~\cite{Nakata2017}.) 

We consider a heterostructure that consists of a heavy-metal film on top of a GQP, as shown in Fig. \ref{fig:device}. While GQPs are typically antiferromagnets, we model the GQP as the bosonic Mott-insulating phase of an easy-plane ferromagnetic insulator for simplicity. A further motivation for doing this is that the theoretical description of this phase is well-developed, mostly as a result of its relevance for cold-atom systems \cite{Fisher1989,Greiner2002, Dries2001}.  Our results will carry over to antiferromagnets since spin currents between normal metals and antiferromagnets are similar in description as those between ferromagnets and normal metals \cite{takei2014,cheng2014}. Our goal is to develop the theory for calculating the spin current from the normal metal into the GQP. Our main result is that this spin current, or, more specifically, the communication channel it opens between the magnetic dynamics and the electronic charge current, gives rise to a spin-Hall magnetoresistance that strongly depends on the magnitude of the external field. 

\textit{Magnon spin-Hall magnetoresistance.---}
We start with a general analysis of how the conductance of a normal metal with spin orbit coupling, e.g. Pt, is modified when it is in contact with magnetic insulator. Due to the presence of spin-orbit coupling the spin Hall effect arises in the metal. The transport in the normal metal is described by \cite{spin_hall}
\begin{align}
\mathbf{j}^{q} &= \dfrac{\sigma}{e}\nabla\mu_q -\dfrac{\sigma ' }{2e}\nabla \times \boldsymbol{\mu}\label{eq:spin_hall1},\\
\dfrac{2e}{\hbar}\mathbf{j}_{n}^{S} &= -\dfrac{\sigma}{2e}\nabla(\hat{n}\cdot\boldsymbol{\mu}) - \dfrac{\sigma ' }{e}\left (\hat{n}\times \nabla\right )\mu_q, \label{eq:spin_hall2}
\end{align}
where $\mathbf{j}^{q}$ and $\mathbf{j}^{S}_{n}$ are the charge current and the spin current (polarized in the $\hat{n}$ direction) respectively, $\sigma$ is the electrical conductivity, $\sigma'$ the spin Hall conductivity, $\mu_q$ the electrochemical potential, and $\boldsymbol{\mu}$ is the spin accumulation. The spin Hall effect leads to a nonzero spin accumulation in the metal which follows the diffusion-relaxation equation $\nabla^2\boldsymbol{\mu}=\boldsymbol{\mu}/l^2_s$, where $l_s$ stands for the spin relaxation length. We consider a charge current flowing in the $y$-direction and the interface in the $yz$ plane (see Fig.~\ref{fig:device}), such that only the $z$-component of the spin polarization is relevant, i.e., $\boldsymbol{\mu}=\mu_z(x)\mathbf{z}$. The spin current transmitted through the interface is polarized in the $z$-direction and we assume that follows a linear-response relation $\mathbf{j}^{S}|_{\text{int}} \cdot {\bf x}=G\mu_z$, where $\mu_z$ is the spin accumulation at the normal-metal side of the interface, and where we have assumed that the magnon chemical potential of the magnetic insulator is zero. (While this is an incorrect assumption for YIG at room temperature \cite{Ludo2016} we expect it to be appropriate for the GQP, because the relaxation of spin is likely to be dominant over spin-conserving relaxation of energy at low temperatures.) We solve Eqs.~(\ref{eq:spin_hall1}$-$\ref{eq:spin_hall2}) together with the spin diffusion equation and obtain a thickness-averaged charge current density
$$\langle j_y^{q} \rangle = \left [1+2\dfrac{l_s}{d_N}\dfrac{\sigma'^{2}}{\sigma^{2}}\left (\dfrac{2e^{2}Gl_{s}+\hbar\sigma}{4e^{2} G l_{s} + \hbar\sigma} \right ) \right ]\sigma E,$$
in the limit $d_N/l_s\gg 1$. Ignoring magnetoresistance of the NM itself, all magnetoresistance arises through the magnetic field dependence of the interface spin conductance $G$. The spin-Hall magnetoresistance observed by Nakayama {\it et al.} \cite{Nakayama2013} is modelled by ignoring thermal fluctuations so that $G=0$ when the magnetic order is aligned with the spin-polarization of the electrons in the NM, and  $G=g_{\uparrow\downarrow}/4\pi$ the perpendicular case, corresponding to the the situation that all spin current is absorbed as a torque on the magnetization\cite{Bender2015}. Here, $g_{\uparrow\downarrow}$ is the real part of the spin-mixing conductance that characterizes the efficiency of the coupling between the electrons spins and the magnetization across the interface \cite{Arne2000}. For our purposes, we consider ${\bf B}$ to be always pointing in the $z$-direction. Then, at nonzero temperature, $G>0$ due to the presence of thermal excitations which correspond e.g. in the simplest case to weakly-interacting magnons. Therefore, $G$ will generically depend on the magnitude of the magnetic field. In proximity to the GQP-phase, a small change in the magnitude of the magnetic field may induce a large gap for the spin excitations which, as we show below, manifests itself in a large field-induced change in $G$ that is observable as spin-Hall magnetoresistance. 
 
\textit{Interface spin current.---}  
We proceed by evaluating the interface spin conductance following the Green's function formalism for spin transport through heterostructures that contain magnetic insulators \cite{Jiansen2017}. We write the spin current across the interface as
\begin{equation}\label{eq:current}
j^{S}=\int\dfrac{d\varepsilon}{2\pi}\ \mathcal{T}(\varepsilon)\left [n_{B}\left (\dfrac{\varepsilon -\mu_z}{k_BT} \right )- n_{B}\left (\dfrac{\varepsilon}{k_BT} \right )\right ],
\end{equation} 
where $n_B(x)=(e^{x}-1)^{-1}$ is the Bose-Einstein distribution function and $\mathcal{T}(\varepsilon)=2\text{Tr}\left [\text{Im}\left (\Sigma^{\text{NM}}\right )\text{Im}\left (\mathcal{G}^{(+)}\right ) \right ]$ is the transmission function at energy $\varepsilon$ that depends on the imaginary parts of the retarded Green's function $\mathcal{G}^{(+)}$ and the self-energy $\Sigma^{\text{NM}}$ of the magnons due to the coupling with the NM. This formula is exact to lowest order in the interface exchange coupling between electron spins and spins $S$ in the magnetic insulator and applies to the case that the magnons are interacting strongly. By expressing the interface spin current in term of the magnon Green's function we assume, however, that magnons are the relevant excitations of the ferromagnet. As the interaction among the electrons in the metal and the spin in the magnet is localized at the interface, the imaginary part of the self-energy is $\text{Im}\left [ \Sigma^{\text{NM}}\right ]_{ij} = -2g_{\uparrow\downarrow}a^{3}(\varepsilon-\mu_z)\delta_{ij}\delta_{ik}/4\pi S$, where $i,j$ label sites of the underlying lattice, $a$ is the lattice constant, and the Kronecker delta $\delta_{ik}$ enforces the self energy to be nonzero only at the interface, with $\{ k \}$ the collection of lattice sites of the magnetic insulator adjacent to the normal metal. 

\begin{figure}[htbp] 
    \centering
    \includegraphics[width=0.48\textwidth]{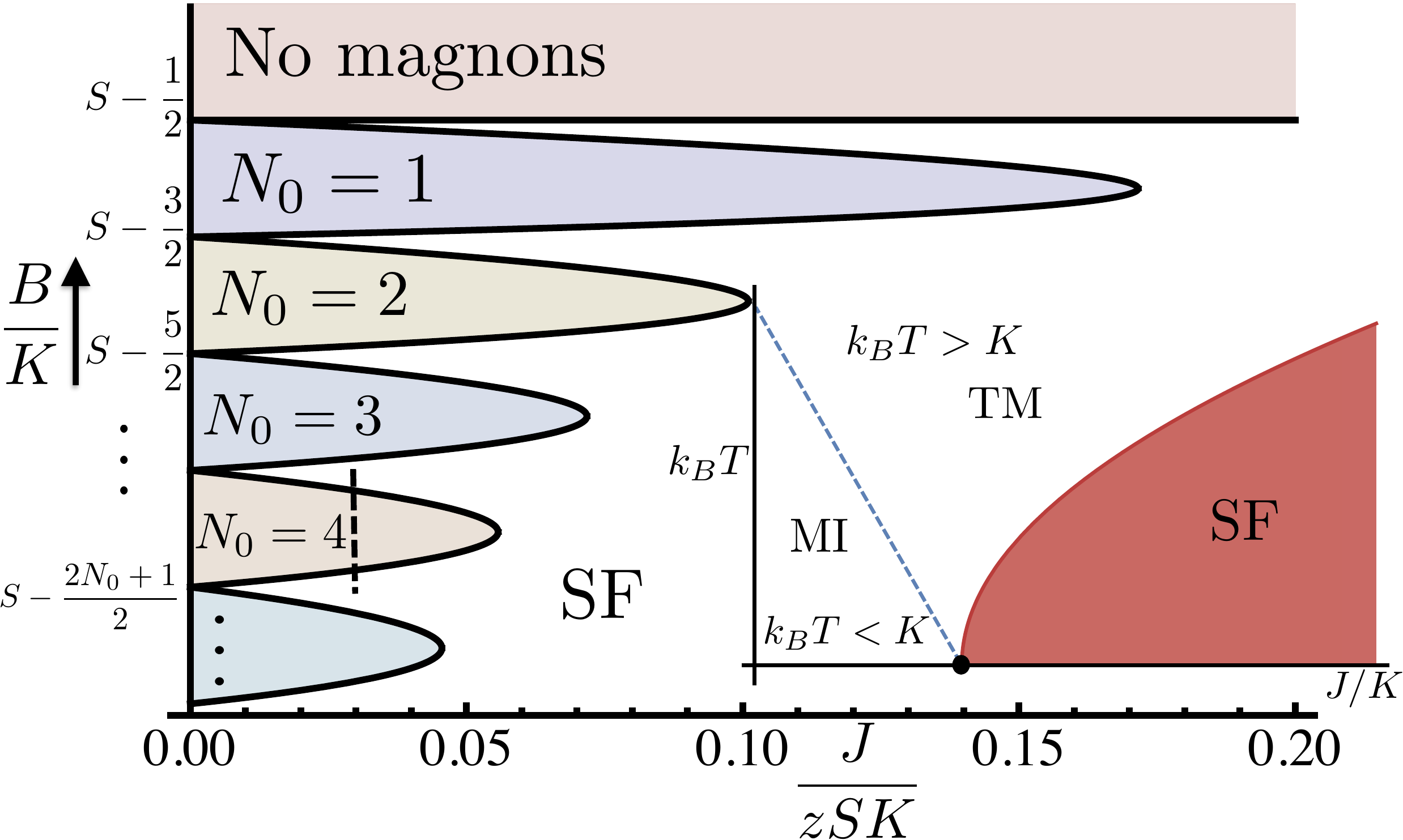}
    \caption{(Color online.) Phase diagram of the Mott insulating (MI) phase obtained from the decoupling approximation elaborated in  \cite{Dries2001}. Each Mott lobe is characterized by an integer occupation $N_0$ of magnons per site. This number can be tuned, for example, by modifying the amplitude of the magnetic field. For nonzero exchange $J$ the lobes are disconnected by the superfluid (SF) phase. The dashed black line shows the values of parameters taken in Fig.~\ref{fig:cond_B}. (Inset) Schematic phase diagram for fixed field. The temperature scale $K/k_B$ is the crossover above which the system behaves as a weakly-interacting gas of magnons in the SF phase or the thermal magnon phase (TM), whereas below this temperature the low-lying excitations are magnonic quasi-particle and quasi-hole excitations.}
    \label{fig:phases}
 \end{figure}
\textit{Mott insulator in the easy-plane ferromagnet.---}
We now consider the magnetic insulator to be an easy plane ferromagnet in presence of a magnetic field $\mathbf{B}=B\mathbf{z}$, described by the Hamiltonian
\begin{equation}\label{eq:01}
\mathcal{H}=-\dfrac{J}{2\hbar^2}\sum_{\left <i,j\right >}\mathbf{S}_{i}\cdot\mathbf{S}_{j}+\dfrac{K}{2\hbar^2}\sum_{i}(S_{i}^{z})^2+\dfrac{B}{\hbar}\sum_{i}S_{i}^{z},
\end{equation}
where $J$ is the exchange interaction among nearest neighbours, and $K$ is the anisotropy. We apply the Holstein-Primakoff transformation 
$\spp_{i}=\hbar a^{\dagger}_{i}\sqrt{2S-a^{\dagger}_{i}a_i}$, $\smm_{i}=(S_{i}^{+})^{\dagger}$, and $ S_i^z=\hbar(a^{\dagger}_{i}a_i-S)$.
With this transformation (in the linear approximation) the Hamiltonian of Eq.~\eqref{eq:01} becomes a Bose-Hubbard Hamiltonian
$$\mathcal{H} =  -\text{t}\sum_{\langle i,j\rangle}a^{\dagger}_{i}a_{j} + \dfrac{U}{2}\sum_{i}n_{i}(n_{i}-1) -\mu\sum_{i}n_{i},$$
with hopping $\text{t}= JS$, effective chemical potential $\mu = -K(1-2S)/2 -B- JSz/2$, and on-site interaction $U=K$. Here, $z$ represents the coordination number.  We assume that $K/J\gg 1$ and $\mu>0$ which leads to the (magnonic) Mott insulating phase for low temperatures \cite{Fisher1989} (see Fig.~\ref{fig:phases} for the phase diagram). For values of the field $B$ that correspond to a commensurate filling $N_0$ of magnons, the system is has a gap $\sim K$ due to interactions. For vanishing exchange interaction, this Mott-insulating state corresponds to a product state $|\Psi\rangle \propto \prod_i  | S, -S+N_0 \rangle_i$, where $|S,m_S \rangle_i$ are the eigenstates of the operator $\hat S_î^z$. For increasing $J/K$, the system undergoes a transition to a gapless \textit{XY} magnet that is spin superfluid, and where the expectation value of the transverse spin is nonzero.

\begin{figure}[htbp] 
    \centering
    \includegraphics[width=0.45\textwidth]{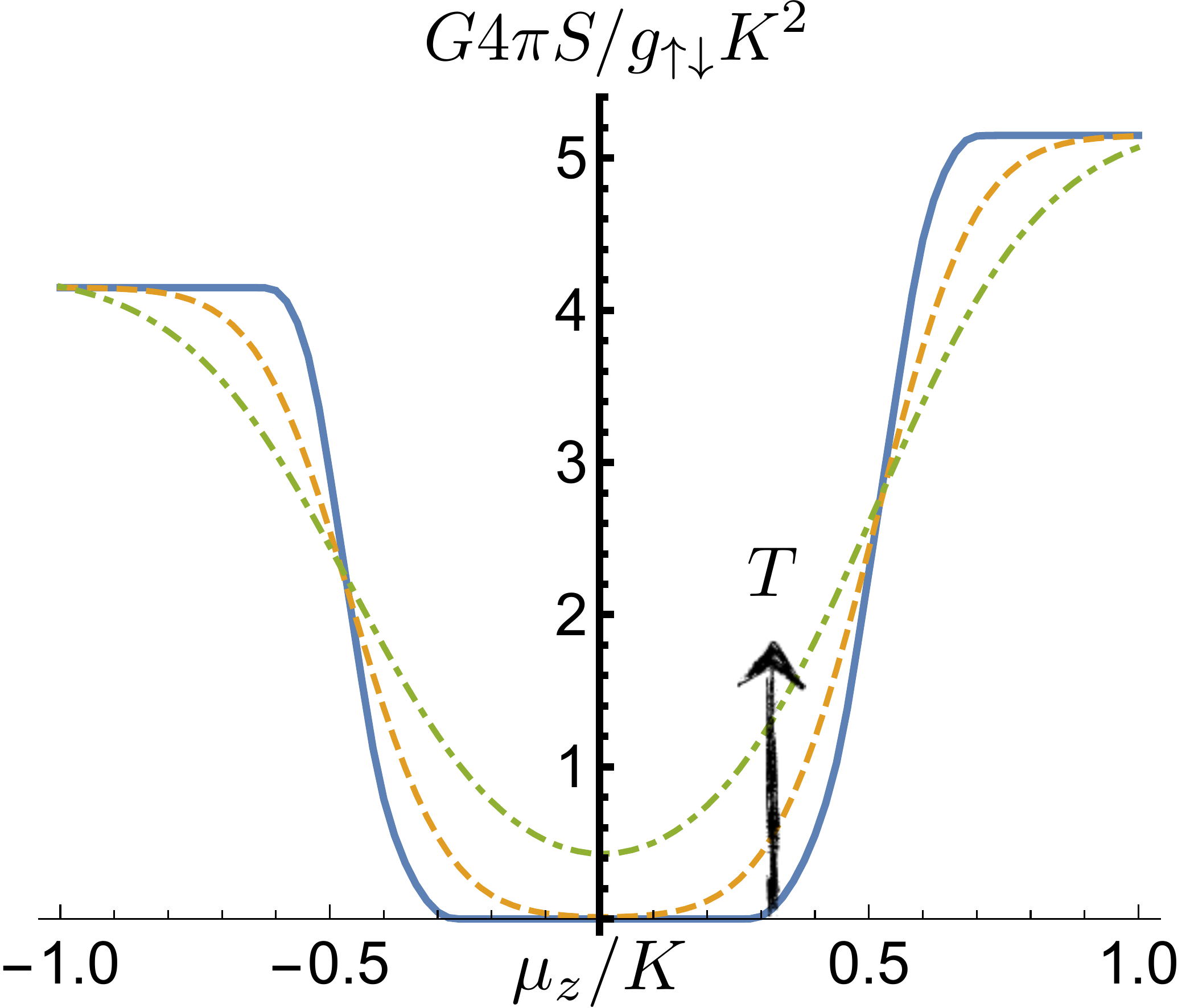} 
    \caption{(Color online.) Interface spin conductance as a function of spin accumulation for different values of the temperature. This calculation was performed considering $N_0=4$, $B/K= S-4$, and $J/zSK=0.05$. The temperatures are $k_BT/K = 0.005$ (solid blue)$,\ k_BT/K =0.05$ (dashed yellow), and $\ k_BT/K =0.1$ (dot-dashed green). }
    \label{fig:conductance_full}
 \end{figure}
In the Mott insulator the low-energy excitations are quasi-particle-hole excitations. Following the mean-field theory developed in \cite{Dries2001} we write the magnon propagator of the magnonic Mott insulator as
\begin{equation*}
\mathcal{G}^{(\pm)}(\mathbf{k},\omega) =\dfrac{Z_{\mathbf{k}}}{\hbar\omega^{\pm}-\varepsilon^{\text{qp}}} + \dfrac{1-Z_{\mathbf{k}}}{\hbar\omega^{\pm}-\varepsilon^{\text{qh}}},
\end{equation*}
were $\hbar\omega^{\pm} = \hbar\omega \pm i|\hbar\omega|\alpha$  with $\alpha$ the bulk Gilbert damping constant, $\varepsilon(\mathbf{k})=-\text{t}\sum_{j}\cos  k_j a$ the dispersion relation of magnons in a cubic lattice of side $a$, $\varepsilon^{\text{qp,qh}}=-\mu +U(2N_0-1)/2 + \left  (\varepsilon(\mathbf{k})\pm\hbar\omega(\mathbf{k})\right  )/2$ the dispersion of quasi-particle(hole) excitations, the energy $\hbar\omega(\mathbf{k})=\sqrt{U^{2}+(4N_0+2)U\varepsilon(\mathbf{k}) + \varepsilon^{2}(\mathbf{k})}$, and $Z_{\mathbf{k}} = \left (U(2N_0+1) + \varepsilon(\mathbf{k}) +\hbar\omega(\mathbf{k})\right )/2\hbar\omega(\mathbf{k})$ the probability of generating a quasi-particle excitation \cite{vanOosten2005}. We consider the small Gilbert damping regime $(\alpha\ll 1)$ that is typical for magnetic insulators, so that Eq.~\eqref{eq:current} becomes
\begin{align}\label{eq:spin_current}
j^{S} = \dfrac{ 
g_{\uparrow\downarrow}}{4\pi S}\int_{\text{\tiny{1BZ}}}&\dfrac{d^3k}{(2\pi)^{3}}\  Z_{k}E^{\text{qp}}_{k}\Delta n_{B}\left (\beta E^{\text{qp}}_{k}\right )\nonumber \\
&+ (1-Z_{k})E^{\text{qh}}_{k}\Delta n_{B}\left (\beta E^{\text{qh}}_{k}\right ),
\end{align}
where $E^{\text{qp,qh}}_{k}=(\varepsilon^{\text{qp,qh}}-\mu_z)$, $\beta=1/k_BT$, and $\Delta n_{B}(x)=n_{B}(x)-n_{B}(x+\beta\mu_z)$.
From Eq.~\eqref{eq:spin_current} we compute the interface spin conductance of the system as ${G}=\partial_{\mu_{z}} j^{S}$. A straightforward calculation leads to
\begin{align}\label{eq:cond_exact}
{G}=&\dfrac{g_{\uparrow\downarrow}}{4\pi S}\int \dfrac{d^{3}k}{(2\pi)^{3}}Z_{\mbf{k}}\left [ F\left (\beta E_{\mbf{k}}^{\text{qp}}\right ) -\Delta n_{B}\left (\beta E_{\mbf{k}}^{\text{qp}}\right )  \right ]\nonumber\\
&+(1-Z_{\mbf{k}})\left [F\left ( \beta E_{\mbf{k}}^{\text{qh}}\right ) -\Delta n_{B}\left ( \beta E_{\mbf{k}}^{\text{qh}}\right )  \right ],
\end{align}
where $F(x)=4x/\sinh^{2}(x/2)$.
 
\textit{Results.---} We numerically evaluate the integral in Eq.~\eqref{eq:cond_exact} as a function of the spin accumulation $\mu_z$ for different values of the temperature (See Fig.~\ref{fig:conductance_full}).
Due to the gap, the conductance exhibits plateaus as a function of $\mu_z$ similar to what is observed for electrons in the Coulomb-blockade regime \cite{coulombblockade}. The value of the conductance for $\mu_z\approx 0$ increases with the temperature. The asymmetry in conductance values for positive and negative large bias derives from the bosonic nature of the magnons. 
\begin{figure}[htbp] 
    \centering
    \includegraphics[width=0.45\textwidth]{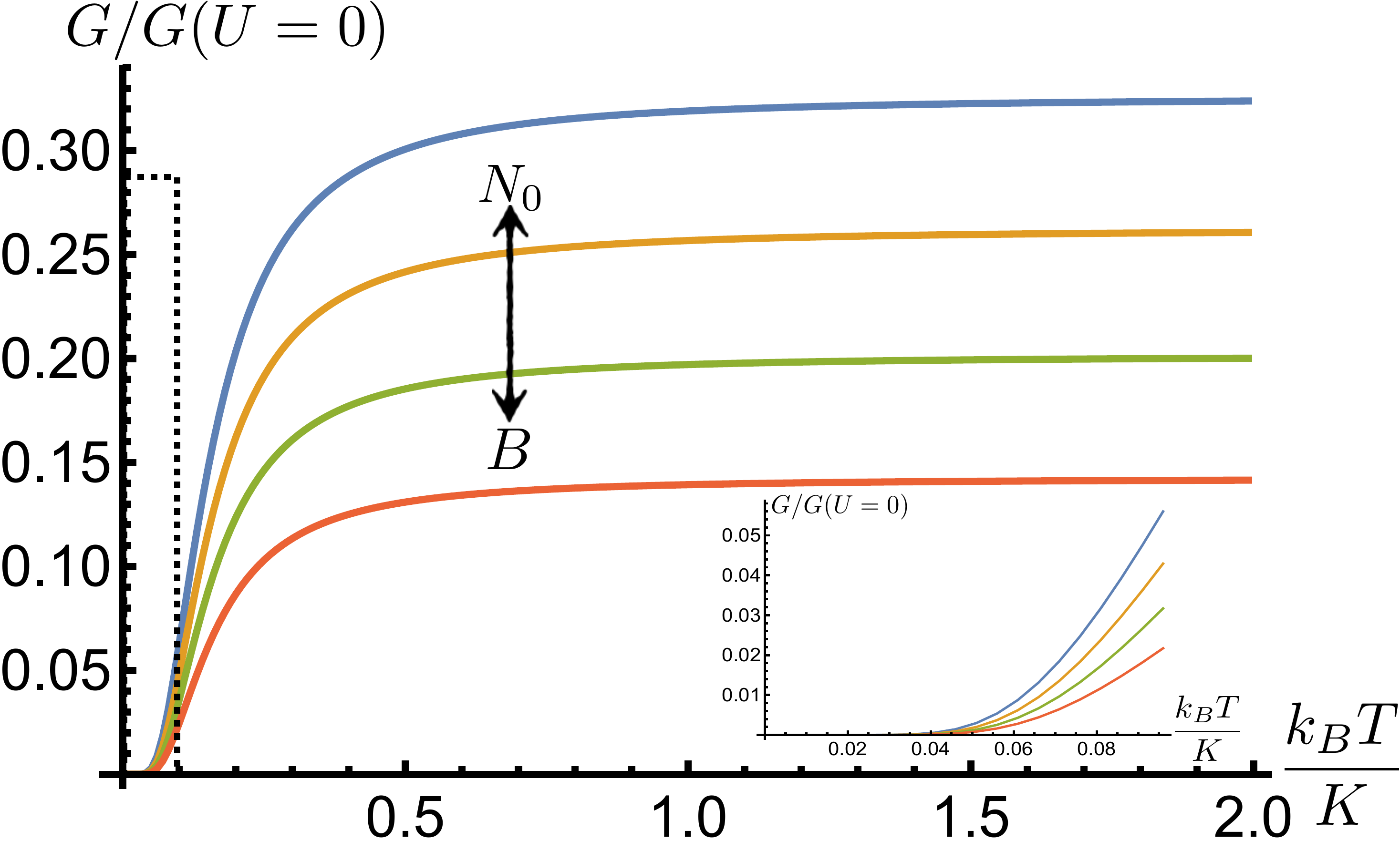} 
    \caption{(Color online.) Temperature dependence of the linearized interface spin conductance normalized to the case of non-interacting magnons, $G(U=0)$ in absence of magnetic field. The different curves represent a state with a different magnon filling fraction. We considered $J/zSK=0.03$, $N_0 = 2$ (orange), $N_0 = 3$ (green), $N_0 = 4$ (yellow), and $N_0 = 5$ (blue), and the magnetic field $B/K= S-N_0$ respectively, which corresponds to the value in the middle of each the Mott lobes. (Inset) Zoom of the normalized conductance behaviour for small temperatures (dashed rectangle in the main plot).}    
    \label{fig:cond_T}
 \end{figure} 
 
In the linear-response regime $(\mu_z\ll K)$ we approximate $\Delta n_{B}(\beta E_\mathbf{k}^{\text{qp, qh}})\approx 4\beta\mu_z/\sinh^{2}(\beta \varepsilon^{\text{qp, qh}}/2)$, so that the current is $j^{S} \approx G\mu_{z}$, where the spin conductance $G$ is given by
\begin{equation}
G=\dfrac{g_{\uparrow\downarrow}}{4\pi S}\int\dfrac{d^{3}k}{(2\pi)^{3}}Z_{\mathbf{k}}F\left( \beta\varepsilon^{\text{qp}}\right ) +(1-Z_{\mathbf{k}})F\left( \beta\varepsilon^{\text{qh}}\right ).
\end{equation}\label{eq:lin_cond}
In Fig.~\ref{fig:cond_T} we show the temperature dependence of this linearized spin conductance. The increase of the conductance with $N_0$ is consistent with the decreasing size of the Mott lobes (shown in Fig.~\ref{fig:phases}).
 
\begin{figure}[htbp] 
    \centering
    \includegraphics[width=0.45\textwidth]{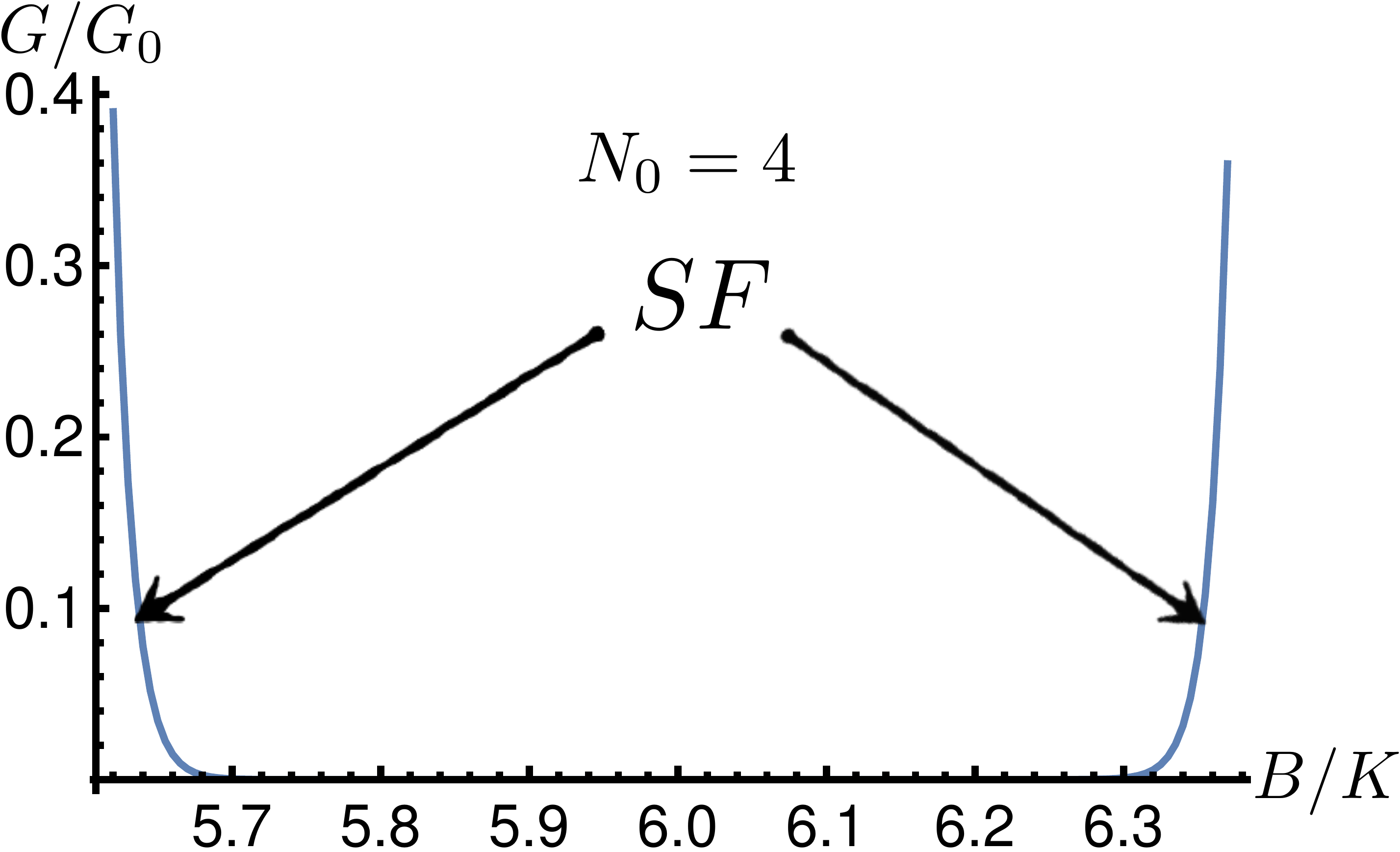} 
    \caption{(Color online.) Linearized interface spin conductance as a function of magnetic field. The magnetic field takes the values shown in the dashed line on Fig. \ref{fig:phases}. As soon as the magnetic field takes values close to the edges of the Mott lobes the conductance increases strongly. These calculations were performed considering $k_BT/K=0.01$, and $J/zSK=0.03$.}
    \label{fig:cond_B}
 \end{figure}

We also consider the dependence of $G$ on the magnitude of the magnetic field (see dashed line in Fig.~\ref{fig:phases}). This allows one to study the conductance of the system upon approaching the MI-SF transition. In Fig.~\ref{fig:cond_B} we show the behaviour of $G/G_0$ as a function of the amplitude of the magnetic field $B$, where $G_0=3g_{\uparrow\downarrow}\zeta(3/2) /16(\pi S)^{5/2}(\beta J)^{3/2}$ is the interface spin conductance between a normal metal and a gas of non-interacting magnons with quadratic dispersion \cite{G_long}. We see that the conductance increases as the magnetic field approaches the boundary of the Mott lobe. Cf. our general discussion of the magnon spin Hall magnetoresistance, this strong change in interface spin conductance modifies the resistance of the adjacent normal metal thereby allowing to probe the phase transition from gapped to gapless magnon state electrically. 

\textit{Discussion.---} We have shown that the interface spin current between a heavy metal and GQP gives rise to magnon spin-Hall magnetoresistance that depends on the magnitude of the field. Assuming that the spin-mixing conductance is of the same order of magnitude as that for a YIG-Pt interface, we expect that the magnitude of this effect is comparable to conventional spin-Hall magnetoresistance \cite{Nakayama2013}. This is because in both cases it relies on the difference between an interfacial spin current that is blocked with one that is fully absorbed.  The spin-Hall magnetoresistance therefore allows one to detect the phase of the quantum magnet electrically. One example of a material that would perhaps be suitable for this purpose is dichloro-tetrakis-thiourea-nickel (DTN) which is an antiferromagnet that shows a GQP state for magnetic fields up to $\sim 1$ Tesla and temperatures of $\sim 1$ Kelvin \cite{Yu2012}. In our study we have taken the temperature of GQP fixed and its spin chemical potential to be zero, considering them to be effectively anchored to a large (phononic) reservoir. Interesting directions for future studies are to step away from this assumption and to consider the internal dynamics of the GQP in response to spin-current injection.  Another interesting situation is a multi-terminal set-up that probes the spin conductivity of the GQP. Shot noise of the injected spin current\cite{Kamra2016} is interesting as this may probe the quenched nature of the number correlations of the GQP. In conclusion, we hope that our work motivates connections between magnon spintronics and quantum magnetism.

\textit{Acknowledgements.---} RD is member of the D-ITP consortium, a program of the Netherlands Organisation for Scientific Research (NWO) that is funded by the Dutch Ministry of Education, Culture and Science (OCW). This work is in part funded by the Stichting voor Fundamenteel Onderzoek der Materie (FOM). RD also acknowledges the support of the European Research Council.

\vspace*{-1cm}
\bibliographystyle{elsarticle-num}

\end{document}